\documentstyle [epsfig,wrapfig,12pt]{article}
\begin{document}
\hfill  ORNL-CCIP-93-14 / RAL-93-069
\vspace{0.5cm}
{
\begin{center}
{\bf SIGNATURES FOR HYBRIDS}
\footnote[1]{This is an expanded version of an
Invited Contribution to the Conference on Exclusive Reactions
at High Momentum Transfers, Marciana Marina, Elba, Italy, 24-26 June 1993.}\\

\vspace{1cm}
T.Barnes\\
Physics Division and Center for Computationally Intensive Physics\\
Oak Ridge National Laboratory, Oak Ridge, TN 37831-6373, USA\\
and\\
Department of Physics and Astronomy, University of Tennessee\\
Knoxville, TN 37996, USA\\

\date{}
\end{center}
}

\begin{abstract}
In this review talk I summarize theoretical expectations for properties of
hybrid mesons and baryons, and discuss the prospects for identifying these
states experimentally.

\end{abstract}

\noindent
{\bf 1. Introduction: Why hybrids rather than glueballs?}
\vskip 0.4cm

Since QCD is a theory which contains both quarks {\it and} gluons as
dynamical degrees of freedom, we would expect to see evidence of both
these building blocks in the spectrum of physical color-singlet hadrons.
There is much indirect evidence of
gluonic basis states in mixing effects, for example in the
Breit-Fermi one-gluon-exchange Hamiltonian
used in potential models and in the $\eta$ and $\eta'$ masses.
It is remarkable, however, that
of the hundreds of hadronic states now known, most can be
described
as states made only of quarks and antiquarks in the nonrelativistic
quark model, and the remaining problematic resonances
show no convincing evidence for states with
dominant gluonic valence components. Reviews of candidate gluonic states
and other unusual hadronic states from the viewpoints of
theorists
\cite{revs}
and experimentalists \cite{revexpt} can be found in the proceedings of
recent meetings on hadron spectroscopy.

{\it A priori} one might expect that a search for gluon
constituents in the
spectrum should concentrate on
dominantly pure states of gluons, since the properties of these
might be expected to differ maximally from quark
and antiquark states. This naive expectation
does not survive detailed investigation. The lightest color-singlet
glueball basis states one can form from transverse gluons are $|gg\rangle$,
and the quantum numbers allowed for these states are $I=0, J^{PC}$=
$0^{\pm +},
2^{\pm +}, 3^{++}, 4^{\pm +},$ and so forth.
Since
$q\bar q$ states can also be made with these quantum numbers,
there is a danger of confusion and one would need
to identify all such $q\bar q$ states in
the mass range anticipated for glueballs.
Probably the most reliable glueball
mass estimates are derived from lattice gauge theory. These QCD
simulations anticipate that the scalar should be the lightest glueball,
with a mass of about 1.4 GeV; other glueball states are expected
to lie near or above 2 GeV \cite{LGT}.
A 1.4 GeV scalar glueball should be evident in the $I=0$
$\pi\pi$ S-wave phase shift, which is well-established experimentally to
about 2 GeV \cite{PSpp}.
This phase shift shows a single narrow state, the
$K\bar K$-molecule candidate $f_0(975)$ \cite{WI},
and in addition only a slowly rising
phase underneath this state. If there are more scalar resonances
coupled to this channel
they are
evidently very broad (a broad $^3P_0$ $q\bar q$ state $f_0(1250)$ is
also expected here and has recently been identified in $\gamma\gamma$
\cite{XB}).
Thus glueball spectroscopy may involve a search
for quite broad resonances with conventional $I=0$
$q\bar q$ quantum numbers, which
could be a very unproductive or at best ambiguous exercise.
Although one can
make $J^{PC}$-exotic states from $|ggg\rangle$ basis
states, which would be much more characteristic experimentally,
these should appear at
rather higher masses. A final problem is that expectations for the
preferred decay modes of glueballs are rather obscure
theoretically, since glueballs naively have flavor-singlet couplings,
but these can be masked by phase space and wavefunction effects.
Thus we are led to ask whether gluons might appear
elsewhere in the spectrum as constituents,
in states which might be more distinct experimentally.

This leads us to the subject of hybrids, which are resonances in which both
quarks and gluons are present in the dominant basis state.
Since a gluon transforms as a color octet, it may
be combined with $q\bar q$ and $qqq$ color-octet quark states to make overall
color-singlet hybrid basis states.
These new basis states will lead to additional resonances beyond
those expected by the $q\bar q$ and $qqq$ quark model assignments;
although physical resonances are linear combinations of conventional and
hybrid basis states, even if the mixing is large we will find more
levels than the quark model alone expects. Actually it is somewhat
misleading to
refer to physical resonances as either quark states or
hybrid states, since this mixing implies that
all physical states have hybrid components.
For simplicity we will use a theorist's definition of
``hybrid" as states which are purely
$|q\bar q g\rangle$ or
$|qqqg\rangle$ before the QCD quark-gluon and
gluon-gluon interactions are introduced,
with the understanding that except for exotics there is
ordinary-hybrid configuration mixing
at some level in physical states.

The detailed predictions for the masses and quantum numbers of hybrid
states depend somewhat on the model
used to study them. Nonetheless, as we shall see, there is general agreement
that these states have characteristic features that make them more attractive
experimentally than the broad isosinglet glueballs, so the hybrid sector
is where we may first see clear evidence of resonances with
large or dominant gluonic
components.

\eject
\newpage
\noindent
{\bf
2. Hybrid mesons; masses and quantum numbers}
\vskip 0.4cm

Both meson and baryon hybrids are anticipated theoretically, since we can form
color-singlet basis states from $q\bar q g$ and $qqqg$.
Although these hybrid basis states mix with ordinary quark-model basis
states such as $|q\bar q\rangle$ and $|qqq\rangle$ to form physical hadrons,
we anticipate more physical states than the quark model alone predicts,
and if the mixing is not large, the dominantly hybrid states
may have unusual properties.

\begin{wrapfigure}{r}{4.0in}
\epsfig{file=elba1.epsm,width=4.0in}
\caption{Light hybrid mesons in the bag model [9].}
\label{fig 1}
\end{wrapfigure}

We first consider hybrid mesons because they have a very attractive
feature:
some $|q\bar q g\rangle$
hybrid basis states have exotic-$J^{PC}$ quantum numbers forbidden
to ordinary $q\bar q$ states,
and for these exotics the mixing and possibility of confusion with
$|q\bar q\rangle$
does not arise.
Identification of dominantly hybrid states or other
unusual hadrons which have quantum numbers accessible to conventional quark
model states
will remain an ambiguous exercise until the quark states
in the relevant mass region are well established.
Identifying all the relevant $q\bar q$ states
in the $\approx 1.5-2.5$ GeV mass region of interest for hybrids
will require considerable experimental effort.
In contrast, identification of a state with
exotic
$J^{PC}$ quantum numbers would at least indicate
that we have found a state beyond the conventional quark model. One could then
determine decay modes and search for other members of a multiplet,
to see whether the new state agrees with expectations for a hybrid or glueball
or perhaps a molecular multiquark state. Hadronic molecules such as the
$K\bar K$ candidates $f_0(975)$ and $a_0(980)$ \cite{WI}
and other possibilities such as vector-vector states \cite{VV}
will complicate the identification of hybrids
somewhat by contributing
non-$q\bar q$ resonances to
the meson spectrum, and these
molecules can also have
exotic quantum numbers. Fortunately, molecules should have rather
characteristic features \cite{molec} such as S-wave meson-meson
quantum numbers and masses not far below the associated two-meson threshold,
so it should be easy to distinguish them from hybrids.

The masses of light hybrid mesons $(q=u,d$ and $s)$ have been estimated using
the MIT bag model \cite{BCD,hybag},
QCD sum rules \cite{hysr},
the
flux tube model \cite{hyft} and
heavy-quark lattice gauge theory \cite{hylgt}.
The flavor quantum numbers of $q\bar q g$ are those of the $q\bar q$ pair,
so light hybrids span conventional flavor nonets.
In the
bag model the lightest gluon mode is TE, with $J^P=1^+$; combining this with
the $J^P=0^-$ and $1^-$ of $q\bar q$ in their lowest bag modes gives
$$
J^{PC_n}(q\bar q g) = \cases{0^{-+},
 1^{-+}, 2^{-+}, & $S_{q\bar q}=1$;\cr
1^{--} &$S_{q\bar q}=0$ \ .\cr }
\eqno(1)
$$
Consideration of other multiplets shows that all $J^{PC}$ can be made from
$q\bar q g$ basis states.
The $J^{PC}=1^{-+}$ is of special interest because this is the lightest
$J^{PC}$-exotic predicted by the bag model. The exact mass depends on the
details of the bag parameters chosen, but
is typically estimated to be about $M(1^{-+})\approx 1.5$ GeV. The
spectrum of
physical hybrids ($q\bar q g$ states mixed perturbatively with
$q\bar q$, $gg$ and $q\bar q gg$ components)
found by Barnes, Close and
deViron \cite{BCD} in this multiplet is shown in Fig.1. Similar results
for the bag model hybrid spectrum were reported by
Barnes and Close, Chanowitz and Sharpe,
and Flensberg, Peterson and Sk\"old
\cite{hybag}.

The predictions of the flux tube
model for hybrids are especially interesting because
this model gives good results for both the spectrum
and decays of conventional quark states. The lightest hybrid flux-tube
multiplet is quite rich, and contains the quantum numbers
$J^{PC} = 1^{\pm\pm}, 2^{\pm\mp},
1^{\pm\mp}$ and $0^{\pm\mp}$, all approximately degenerate. Note the
presence of the $1^{-+}$ exotic,
as in the bag model, and the additional exotics $0^{+-}$ and
$2^{+-}$. The mass of this multiplet
(with $q=u,d$)
is somewhat higher than bag model
expectations, and has varied between $\approx 1.7$
GeV and 2.0 GeV in the flux tube
literature \cite{hyft}.

QCD sum rules can be
used to estimate the masses of the exotic hybrids, since these are expected
to be the lightest resonances in these $J^{PC}$-exotic channels.
There have apparently been several algebraic errors in the sum rule literature
in the past, which are discussed by Latorre, Narison, Pascual and Tarrach
\cite{LNPT}. The most recent work of Latorre {\it et al.} \cite{LPN} finds
$q=u,d$ exotic hybrid masses
of $M(1^{-+})\approx 2.1$ GeV and $M(0^{--})\approx 3.8$
GeV.

Finally, Wilson loop techniques have been applied to the study of hybrids
in the heavy-quark limit by the Liverpool group
\cite{hylgt}.
Their result for the lightest ``E$_u$" hybrid multiplet, which
has a $1^{-+}$ member and other exotics,
is $M(hybrid)\approx m_{Q\bar Q}$ + 1 GeV. If applicable to
light quarks, this result is broadly consistent with the estimates found using
other models, except perhaps the higher QCD sum rule estimate.

Of course, one may also attempt to identify non-exotic hybrids in the
meson spectrum, and there are many suggestions for possible hybrid states
in the literature (see for example \cite{otherh},
the reviews \cite{revs,revexpt}
and the other phenomenological references in the bibliography).
These
attempts will probably remain controversial due to confusion with
$q\bar q$ states
until true exotic hybrids are identified,
following which the identification of non-exotic partners in hybrid multiplets
should be more straightforward.

There has recently been considerable interest in searches for heavy-quark
hybrids. The advantage of these systems is that
they should be relatively pure in Hilbert space, because the
mixing between
$|q\bar q g\rangle$ and other basis states is driven by
$\vec \jmath^{\; a}\cdot\vec A^a$,
which is reduced due to the lower velocities of heavy quarks.
Since the
spectrum of heavy quarkonium is less complicated than in
light hadronic systems, experimental identification of heavy hybrids
might
be more straightforward.

There are several model calculations of heavy-quark hybrid masses in the
literature.
The mass estimated
for the lowest $c\bar c$-hybrid multiplet has varied over
the range $\approx 4.2-4.5$ GeV in flux tube references \cite{hyft}.
Perantonis and Michael
\cite{hylgt} find 4.04 GeV for the lightest $c\bar c$-hybrids
in quenched heavy-quark
lattice gauge theory,
and estimate 4.19 GeV as a full QCD result. They also note that the effective
$Q\bar Q$ potential for hybrids is rather shallow, so
radially and orbitally excited hybrids should lie not far above
the hybrid ground state.
Finally, Narison \cite{hysr}
quotes a QCD sum rule result of 4.1 GeV for the $1^{-+}$ exotic
$c\bar c$-hybrid, consistent with lattice gauge
theory and the lower flux-tube results.
Theoretical estimates
of the mass of the lightest $c\bar c$-hybrid multiplet
are thus typically about $4.2 \pm 0.2$ GeV.
Typical mass estimates for $b\bar b$-hybrids are 10.5 GeV
(Narison, sum rules \cite{hysr})
and up to 11.1 GeV (Merlin
and Paton, flux tube model \cite{hyft}).

In summary, although
there is considerable variation in detail, all these approaches
predict
that the
lightest exotic hybrid mesons have masses of $\approx 1.5 - 2.0$ GeV, and in
heavy-quark systems (from the flux-tube model and lattice gauge theory)
lie near the lightest $Q\bar Q$ mass plus $\approx 1-1{1\over 2}$ GeV.
The exotic quantum
numbers $J^{PC}=1^{-+}$ are often suggested for experimental searches
in light-quark systems,
because all techniques find a hybrid with these quantum numbers, and
the flux tube model predicts
in particular that the $1^{-+}$ state with isospin 1
should be relatively narrow.

\vskip 0.6cm
\noindent
{\bf
3. Signatures for hybrid mesons: decays and couplings}
\vskip 0.4cm

In addition to general searches for extra or $J^{PC}$-exotic resonances,
one can use theoretical expectations for hybrid
decay modes to motivate experimental
searches in particular strong final states or
in electromagnetic processes.

Theoretical models predict rather characteristic
two-body decay modes for hybrids.
Both
flux tube \cite{hyft} and
constituent gluon \cite{hycg}
models
find that the lightest hybrids decay preferentially
to pairs of one $L_{q\bar q}$=0
and one $L_{q\bar q}$=1 meson,
for example $\pi f_1$ and $\pi b_1$. These unusual modes
have received little
experimental attention because they involve complicated final states,
which may explain why hybrids were not
been discovered
previously.
There is already
some data on these
final states; an $I=1$, $J^{PC}=1^{-+}$ exotic is reported at 1.775 GeV
by a
SLAC photoproduction experiment \cite{Condo}, and will be studied by E687
at Fermilab. The Crystal Barrel collaboration has results for
$\pi b_1$ but sees no evidence for unusual states \cite{Doser}.
Several experiments plan future studies of these
channels, including E818 (to study $\pi^- f_1$ ) \cite{Chung}
and E852 (to study $\pi f_1$ and $\pi \eta$) \cite{ad}, both at BNL.
Finally, E781 at Fermilab plans a sensitive search for
hybrids using the Primakov effect
\cite{Ferbel}.

Much of the experimental work on possible hybrids
has concentrated on the $\pi\eta$ system, since this is easy to analyze
(all odd-L $\pi\eta$
waves are exotic) and an
important P-wave contribution in the angular distribution near and above
1.3 GeV has long been known \cite{apel}.
Of course the important question is whether this
P-wave amplitude is resonant, and the analysis
is complicated by the interference of the
P-wave with the resonant D-wave $a_2(1320)$ amplitude.
There are
recent experimental studies
of this final state
by GAMS \cite{GAMS} and E179 at KEK \cite{KEK} which suggest that
the exotic P-wave
amplitude is indeed resonant.
The VES group has reported results from
phase shift analyses of
$\pi^-\eta$
and
$\pi^-\eta'$, and see evidence for
a broad enhancement at 1.6 GeV
which appears more strongly in
$\pi\eta'$
\cite{VES}. They suggest that this may be due to a $q\bar q g$ exotic
state (perhaps not resonant)
because this behavior was predicted by Close and Lipkin \cite{CLip}.
The Crystal Barrel collaboration
reports a nonresonant or at least very broad
P-wave \cite{Doser}.
Several other experiments have reported
results for $\pi\eta$ and related channels such
as $\pi\eta'$; see for example the HADRON93 proceedings for summaries
of recent work.
In view of the disagreements between experiments and
complications in the analyses
the nature of this P-wave amplitude
should probably be considered an open
question until the phase motion of the P-wave has been accurately
understood.

Heavy-quark hybrids, in particular
charmonium hybrids, may be accessible at high-luminosity $e^+e^-$
machines.
One may search for the ``extra" $1^{--}$ hybrid states directly
through a high-statistics scan of R; the
hybrids are expected
to appear relatively weakly since they must couple
to the photon through their
$c\bar c$ components,
but if the hybrids lie below their preferred S+P
open-charm decay threshold at $\approx 4.3$ GeV they may appear as
relatively narrow peaks.
It should be possible to produce charm hybrids with other quantum numbers
than $1^{--}$ through cascade decays, in which an initial high-mass
$1^{--}$ $c\bar c$
pair cascades to a charm hybrid plus a light hadronic
system such as $\eta$ or $\pi\pi$.
Flux tube arguments suggest that the charm hybrid and light hadronic
system should be produced in a relative P-wave, which may
assist in the identification
of hybrid states with specified quantum numbers. The prospects for detecting
heavy-quark hybrids using these techniques have recently been discussed
by Barnes and Close \cite{tcf}.

\newpage
\noindent
{\bf
4. Hybrid baryons}
\vskip 0.4cm

One may also form color-singlet quark-and-gluon basis states from
$qqqg$, and these are expected to lead to
hybrid
baryon resonances not predicted by the naive $qqq$ quark model.
There are no hybrid baryon exotics
since all $J^P$ can be made from $qqq$ quark model
states, so one must
identify hybrid baryons as additional states
with conventional baryon quantum numbers in a well
established $qqq$ background.

\begin{wrapfigure}{r}{4.0in}
\epsfig{file=el2.epsm,width=4.0in}
\caption{Bag model spectrum of light nonstrange hybrid baryons [29].}
\label{fig 2}
\end{wrapfigure}

To date
detailed calculations of
the masses of hybrid baryons have been reported only in
the bag model.
In the bag model the color-octet $qqq$ components
of the lowest unmixed $|qqqg\rangle$ hybrid baryon
basis states
span a {\bf 70} under SU(6), which has the spin-flavor decomposition
(in $^{2j+1}D(SU(3)_f)$ notation) ${\bf 70} =
{}^2{\bf 10}\oplus
{}^4{\bf 8}\oplus
{}^2{\bf 8}\oplus
{}^2{\bf 1}$. Since a transverse gluon
contributes only ${}^3{\bf 1}$ states,
the final $|qqqg\rangle$ hybrid basis states do not
span the complete sets of spin-flavor quantum numbers we
associate with irreducible SU(6)
multiplets. The spin-flavor SU(2)$\otimes$SU(3)
decomposition of
the lightest multiplet of $|qqqg\rangle$ basis states
for $q=u,d,s$
combined with a spin-1 gluon
is
$$ qqqg = \ \
^4{\bf 10}\oplus \
^2{\bf 10}\oplus \
^6{\bf 8}\oplus  \
( ^4{\bf 8})^2 \oplus \
( ^2{\bf 8})^2 \oplus \
^4{\bf 1}\oplus \
^2{\bf 1} \ ,
\eqno(2)
$$
so we expect this set of light hybrid baryon states in addition to the
complete $qqq$ SU(6) multiplets.
The lightest gluon mode has $J^P=1^+$ in the bag model,
which therefore predicts that all these lightest hybrid baryons
have $(+)$-parity.

The spectrum of light, physical
(mixed with $qqq$ and $qqqgg$)
hybrid levels
with $q=u,d$
found
by Barnes and Close \cite{bc} is shown in Fig.2. A very similar
pattern of multiplet splittings was
found by Golowich, Haqq and Karl \cite{GHK}, albeit with an overall
mass scale about 200 MeV lower, due to their choice of more conventional
bag model parameters. This work was extended to $q=u,d,s$ by Carlson
and Hansson \cite{CH}.

Since the nonstrange baryon spectrum is
reasonably well established to about 2
GeV, one might expect to confirm or refute the bag model description
of the hybrid baryon spectrum easily. Remarkably this has not
been possible, because there
are $qqq$ experimental candidates already known near each of the light
hybrid levels predicted. This may imply hidden hybrid levels near the
dominantly $qqq$ levels, or perhaps some resonances usually assigned to
$qqq$ are actually hybrids. The Roper is often cited as a possible
misidentified hybrid \cite{ZP,CL},
because the lightest hybrid baryon in the bag model
has Roper quantum numbers, and Golowich, Haqq and Karl
\cite{GHK} predicted a mass
for this hybrid of about 1400-1450 MeV, consistent with the Roper mass.
Note that the expectation of a hybrid baryon near the Roper mass is
closely linked to the bag prediction of a $1^{-+}$ exotic hybrid meson
near $1{1\over 2}$
GeV, since these predictions use similar bag model parameters.
The flux tube model
probably supports a
radial-$qqq$ assignment
for the Roper because flux tube hybrid baryons
are expected to
occur at much higher masses \cite{NIft},
although detailed spectrum calculations in this model have not
yet been carried out.

Although we cannot distinguish between dominantly $qqq$ and $qqqg$ baryons
by their quantum numbers,
one might identify a dominantly-$qqqg$ (or excited flux tube)
hybrid baryon by
anomalous
couplings relative to conventional $qqq$ baryons;
possibilities include both strong and electromagnetic couplings.
There are indications that some
familiar predictions for light baryon properties
in the $qqq$ quark model are rather insensitive
to admixtures of $qqqg$ basis states
\cite{LK}, and this mixing may be large.
If these results apply throughout the spectrum
and the mixing angles are indeed large,
the distinction between ordinary and hybrid baryons
will be lost,
and the only evidence for hybrids would be an overpopulation
of levels relative to the $qqq$ quark model.
Fortunately
the success of the unmixed $qqq$ quark model in describing
the properties of experimental baryon resonances makes
this complicated
strong-mixing scenario seem unlikely.

First consider hybrid baryon strong decays.
The two-body decay modes of hybrid baryons in the flux tube model
should satisfy a selection rule similar to that for hybrid mesons. In the
baryon case the hybrid flux tube basis state
has a spatially-odd wavefunction for reflection of the
flux tube through the $qqq$ plane, at least in the
heavy-quark
limit. When a $q\bar q$ pair is formed through flux tube breaking, with
$qqq$ and $q\bar q$ final states, the initial odd
spatial symmetry will lead to
a small matrix element unless the final meson or baryon has an
internal orbital
excitation. Since P-wave $q\bar q$ mesons have masses of $M\geq 1.2$ GeV,
the preferred hybrid baryon
decay modes will probably be a P-wave baryon plus a
pion, for example N(1520)$\pi$. Since these modes are closed to the N(1440)
Roper, it is surprising that the Roper is so broad if it is indeed
a hybrid.

One may also use electromagnetic couplings to search for hybrid baryons and
to test quark model predictions for conventional $qqq$ assignments. This
approach will be especially attractive in the future because
photoproduction and electroproduction
amplitudes will be measured with greatly improved accuracy at
CEBAF. Barnes and Close \cite{bc2} studied the photoproduction amplitudes
of the light hybrid multiplet in Fig.2, and found a very characteristic
selection rule; the photoproduction amplitudes of
some of the light hybrids, those with $^4{\bf 8}$ $qqq$ substates
in their
$qqqg$ component,
vanish from proton
but not neutron targets. This is a generalization of an excited-$qqq$ selection
rule previously found by Moorhouse \cite{Moor}, and applies to the lightest
hybrid baryon, which is the candidate Roper state.
Since the Roper does not satisfy
this selection rule
experimentally, the bag model hybrid baryon does not appear to be a
good description of the Roper. Alternatively it has been suggested
that the Roper may be a different combination of
hybrid basis states than these
simple bag model calculations find \cite{ZP}.
To distinguish between hybrid and excited-$qqq$ assignments for states
such as the Roper it will be useful to have accurate quark model
predictions for photoproduction and electroproduction amplitudes, since the
$qqq$ quark model wavefunctions are reasonably well established.
At moderate $Q^2$, early results
by Close and Li \cite{CL}
suggested that the radial-$qqq$ assignment disagreed with
existing Roper electroproduction data \cite{Stoler},
but quark model calculations
by Warns, Pfeil and Rollnik \cite{WPR} and Capstick \cite{Simon} found that
more accurate wavefunctions alter these results, and lead to
smaller radial-$qqq$ electroproduction amplitudes.
More recently, Capstick and Keister \cite{SimonBrad} have found that
relativistic effects are quite important in radial-$qqq$ electroproduction,
so previous nonrelativistic calculations
may be inaccurate. Evidently
careful theoretical studies of electroproduction amplitudes are required
for comparison with the accurate experimental data
expected from CEBAF.
In the higher-$Q^2$ regime, where we expect to see increasingly important
contributions from perturbative QCD processes, it may also be possible to
distinguish between dominantly $qqq$ and $qqqg$ baryons.
Carleson and Mukhopadhyay \cite{CM} note that the
transverse electroproduction form factor should distinguish between
these cases, and expect a leading power of
$G_+(Q^2)\propto 1/Q^3$ for a $qqq$ state but
$G_+(Q^2)\propto 1/Q^5$ for electroproduction of a $qqqg$ state.
To test these various predictions it will be important to
measure resonance
electroproduction functions over a wide range of $Q^2$.

\vskip 0.6cm
\noindent
{\bf
5. Summary and Conclusions}
\vskip 0.4cm

In this review we have discussed theoretical expectations for the properties
of hybrid mesons and baryons. Hybrids are hypothetical resonances in which
the dominant basis states incorporate both quark and gluonic excitations.
Although mixing between
pure quark basis states and excited-glue basis
states is anticipated and may be large,
the additional gluonic basis states
will nonetheless lead to more resonances than the quark model
alone anticipates.
Complications due to mixing with quark states can be avoided in
the meson sector, since one can form $J^{PC}$-exotic
combinations such as $1^{-+}$ from $q\bar q g$ basis states. Experimental
searches for hybrids with these or other exotic quantum numbers offer the
best prospects for identifying hybrids, since these cannot be confused with
conventional $q\bar q$ states. We
also discussed predictions for the spectrum
of hybrid mesons and their expected
strong decay modes; the favored pair-production
decay mode for light hybrids is
one L=0 and one L=1 $q\bar q$ meson, and this unusual final state
may explain why hybrids have not yet been reported.
Heavy-quark hybrid mesons may be observable at $e^+e^-$ machines as
small peaks in R
if they lie below this S+P threshold,
or in hadronic cascade decays from initial $Q\bar Q$
states.
Finally we discussed light-quark hybrid baryons and the
prospects for detecting these experimentally,
through studies of the spectrum and decays of nonstrange baryons and their
photoproduction and
electroproduction amplitudes.

\vskip 0.6cm
\noindent
{\bf
6. Acknowledgements}
\vskip 0.4cm

I would like to thank C.E.Carlson and the organizers of the Elba Conference
on Exclusive Reactions at High Momentum Transfers for their kind invitation
to attend this meeting and for the opportunity to discuss the status of
hybrids and related topics with my fellow participants.
I would also like to thank
D.Alde, D.V.Bugg, S.Capstick, S.U.Chung, F.E.Close, A.Dzierba,
D.Hertzog, N.Isgur, Z.P.Li,
J.Paton and E.S.Swanson for
additional discussions of material presented here.
This research was sponsored in
part by the United States Department of Energy under contract
DE-AC05-840R21400, managed by
Martin Marietta Energy Systems, Inc, and by the United Kingdom Science Research
Council through a Visiting Scientist grant at Rutherford Appleton Laboratory.

\end{document}